\documentclass[%
 reprint,
 amsmath,amssymb,
 aps,
]{revtex4-1}

\usepackage{graphicx}% Include figure files
\usepackage{dcolumn}% Align table columns on decimal point
\usepackage{bm}% bold math
\begin{document}

\preprint{APS/123-QED}

%\title{Tensor Non-Gaussianities from EFT: Templates and Detectability}
%\title{Can We Probe Primordial Gravitational Waves by  Non-Gaussianities?}
%\title{Probing Primordial Gravitational Waves by  Non-Gaussianities}
%\title{Probing Primordial Gravitational Waves by  Non-Gaussianities: Templates and Detectability}
\title{Non-Gaussian Features of  Primordial Gravitational Waves}
%\title{Proposing Templates of Tensor Non-Gaussianities for Future Missions}
% Force line breaks with \\

\author{Abhishek Naskar}
\email{abhiatrkmrc@gmail.com}
\author{Supratik Pal}%
 \email{supratik@isical.ac.in}
\affiliation{%
 Physics and Applied Mathematics Unit,
 Indian Statistical Institute, 
 %203 B.T. Road, 
 Kolkata-700108, India
}%

\date{\today}% It is always \today, today,

\begin{abstract}
We  explore possible  non-Gaussian features of primordial gravitational waves   
by constructing  model-independent templates for  nonlinearity parameters of tensor bispectrum.
 Our analysis is based  on Effective Field Theory of inflation that relies on no particular model as such and 
 thus the results are quite generic.  
 The analysis further reveals that chances of detecting squeezed limit tensor bispectrum are fairly higher than 
  equilateral limit.
 We also discuss prospects of detectability in upcoming CMB missions.
 
%\begin{description}
%use the optional argument of the \verb+\item+ command to give the category %of each item. 
%\end{description}
\end{abstract}

\maketitle

%\tableofcontents

   \section{Introduction}
 %{\it Introduction} -- 
 With the recent detection of gravitational waves  from binary mergers 
 by LIGO-Virgo collaboration \cite{ligo1, ligo2, ligo3},
 this field of research has emerged as more exciting than ever.
 In it searches for  primordial gravitational waves (GW henceforth)
 have  found their relevance afresh. Primordial
 GW is generically sourced by inflation and it can serve as a missing link between
 early universe cosmology and its signatures on Cosmic Microwave Background (CMB) observations
 via
 tensor fluctuations. Apart from  finding out primordial features of GW, which
 itself is quite appealing, the detection of tensor modes can also tell us about the
    energy scale of inflation. Nevertheless, it can put inflationary paradigm
    on a firm footing (or put it to tension the other way round).
 However, in spite of painstaking searches for the last few years, no signature of
 tensor modes via its two point correlation function has been observed.
Latest observation from Planck satellite   puts an upper
   bound on the tensor two point correlation, viz., 
   tensor-to-scalar ratio as $r\leq0.07$ at $95\%$ CL \cite{planck} (further improved to
   $r\leq0.064$ in the latest 2018 data release
   \cite{Akrami:2018odb}).  Hence, searching for 
    other characteristics of tensor perturbations, such as the three point function,
    which may have the potential
    to comment on primordial GW, is of extreme relevance  these days. 
    
   Along with the above constraints, Planck \cite{planck,planck2} also set a constraint
     on tensor non-Gaussianities as $f_{NL}^{T} = 400 \pm 1500$ at $68 \%$ CL. 
     Similar constraints can also be found in  \cite{Shiraishi:2014ila}  based on WMAP. 
     These early analyses show that the tensor perturbations can, in principle, significantly deviate  
     from Gaussian nature.
     With future CMB missions like COrE \cite{COrE},  LiteBIRD \cite{Matsumura:2013aja, Suzuki:2018cuy},
     CMB-S4 \cite{Abazajian:2016yjj}, PRISM  \cite{Andre:2013nfa}, PIXIE  \cite{Kogut:2011xw} etc. 
     chipping in,
     it is high time the community  have in their hand couple of useful, model-independent
     templates for tensor non-Gaussianities that can serve as a probe for primordial GW in
     these upcoming missions. This will  help us analyze the prospects of
     detection  and compare with the sensitivity of the upcoming missions.
     Nevertheless, the non-Gaussian signatures may serve as  additional probes of  tensor modes, 
      along with the usual
     tensor to scalar ratio.
     It is worth mentioning that some of the future missions like LiteBIRD \cite{Matsumura:2013aja}
     already aim at detecting 
     tensor non-Gaussianities at 3-$\sigma$. Recently in \cite{Bartolo:2018qqn}
     a formalism is developed to detect non-Gaussianity of tensor sector from
     correlation of three signals in Laser Interferometer Space Antenna (LISA).
     So it is quite  timely
     to explore the three point tensor correlation functions that can open up
     an altogether new direction towards investigation for 
      primordial GW.
     
     In the last couple of years, there has been
     some  progress in this direction. To mention a few, in
      \cite{mald1,mald2} tensor
     bispectrum  is calculated for general single field slow roll inflationary model. 
     Parity violating tensor non-Gaussianities have also been calculated in \cite{soda}.
      These works are extended for generalized G-inflation  
      that takes into account the most general second-order equations of motion 
      for single field models \cite{yamaguchi, {Kobayashi:2011nu}}.
       In \cite{sorbo} a 
      large tensor non-Gaussianity and a nearly Gaussian scalar fluctuation is produced using a pseudo-scalar; 
      and in \cite{aniket} large non-Gaussianity is produced using the coupling between an axion field 
      and a $SU(2)$ gauge field. However, all of them are model-dependent approaches that rely either
      on particular (class of)  models or on specific mechanisms.
  
  In this article we propose a  model-independent framework based on Effective Field Theory (EFT) of inflation developed 
  in \cite{crem2} and derive the most 
  general model-independent third order action for tensor perturbation.
 We compute 
  the generic tensor bispectrum therefrom.  Finally, we search for possible templates for tensor bispectrum
  and comment on the prospects of detectability in upcoming CMB missions.
   
    %%%%%%%%%%%%%%%%%%%%%%%%%%%%%%%%%%%%%%%%%%%%%%%%%%%%%%%%%%%%%%%%%%%%%%%%%%
	\section{Graviton Lagrangian from EFT}
	%{\it Graviton Lagrangian from EFT} --
	In order to propose a generic template  in
	this model-independent analysis, the  graviton Lagrangian has to have 
	the following properties: (i) it has to be more or less generic
	with no {\it a priori} dependence of particular inflation models or particular mechanism to generate
	tensor non-Gaussianities, (ii) it has to be consistently derived
	from EFT and should finally have least number of free parameters. 
	To construct such a generic Lagrangian for the graviton,
	we make use of 
	 the EFT approach
	 developed in \cite{crem2,weinberg}. 
	 In this approach, the inflaton field $\phi$ is a scalar under all diffeomorphisms but $\delta \phi$ breaks the time 
	 diffeomorphism. Using this symmetry of the system and unitary gauge where $\delta \phi=0$, the most general Lagrangian 
	 can be written as
	\begin{widetext}	
\begin{multline}\label{eq1}
	\mathcal{S}=\int d^4x \sqrt{-g}\left[\frac{1}{2}M_{pl}^2R-\Lambda(t)-c(t)g^{00}+
	 \frac{1}{2}M_2(t)^4(g^{00}+1)^2-\frac{\bar{M}_1(t)^3}{2}(g^{00}+1)\delta K_{\mu}^{\mu}
           -\frac{\bar{M}_2(t)^2}{2}\delta K_{\mu}^{\mu2} \right. \\
           \left. 
           -\frac{\bar{M}_3(t)^2}{2}\delta K_{\mu}^{\nu} K_{\nu}^{\mu}\right. 
            \left. +  \frac{M_3(t)^4}{3!}(g^{00}+1)^3-\frac{\bar{M}_4(t)^3}{3!}(g^{00}+1)^2\delta K_{\mu}^{\mu}
            -\frac{\bar{M}_5(t)^2}{3!}(g^{00}+1)\delta K_{\mu}^{\mu 2}\right.\\
            \left.-\frac{\bar{M}_6(t)^2}{3!}(g^{00}+1)
            \delta K_{\mu}^{\nu } \delta K_{\nu}^{\mu}
            -\frac{\bar{M}_7(t)}{3!}\delta K_{\mu}^{\mu 3}
            -\frac{\bar{M}_8(t)}{3!}\delta K_{\mu}^{\mu} \delta K_{\nu}^{\rho} \delta K_{\rho}^{\nu}-
            \frac{\bar{M}_9(t)}{3!}\delta K_{\mu}^{\nu} \delta K_{\nu}^{\rho} \delta K_{\rho}^{\mu}+....\right] 
	\end{multline}
\end{widetext}	
%The dots at the end of the Lagrangian represent higher order terms.  
 Since we are primarily interested
   in three point correlation function, in above Lagrangian \eqref{eq1} terms upto third order in gravitation  
   have been retained. Let us briefly explain the terms in the above action.
   The first three terms are linear in perturbation and the coefficients $c(t)$ and $\Lambda(t)$ 
   can be fixed by background FRW solution. From the equation of motion, one can show that
   $H^2=\frac{1}{3M_{pl}^2}[c(t)+\Lambda(t)]$ and 
$ H^2 + \dot H=- \frac{1}{3M_{pl}^2}[2c(t)- \Lambda(t)]$ representing the background evolution.
Also, for  slow-roll inflation one can readily recast the parameters as $\dot{\phi}(t)^2=2c(t)$, $V(\phi(t))=\Lambda(t)$;
and  get back the usual model-dependent inflationary setup.
The rest of the terms represent effect of perturbations upto third order.
   The higher order operators with coefficients $M_i(t)$ and $\bar{M}_i(t)$ represent different inflation models
   with corresponding perturbations.
   Here $\delta K_{\mu \nu}=K_{\mu \nu}-H h_{\mu \nu}$ is the fluctuation of extrinsic curvature with $h_{\mu \nu}$ is the induced metric.
   So, the above action is a generic one that can represent most of the inflationary models with corresponding perturbations.
One should keep in mind that this is purely gravitational Lagrangian and the scalar perturbation is not explicit but can be reintroduced using 
	 $St\ddot{u}ckleberg$ trick. So the system has three degrees of freedom: one scalar and two tensor.  
	 %In general the
	  %coefficients of the higher order operators have time dependence but we can neglect the dependence as if we
	   %expand in the Taylor series they will be slow roll suppressed. 
	   Any time-dependence on the parameters is slow roll suppressed.

 %\subsection{Graviton Lagrangian}
 
    To work with tensor perturbations  we express the perturbed metric in unitary gauge
  \begin{equation}
  g_{ij}=a^2(t)[(1+2\zeta(t,x)\delta_{ij})+\gamma_{ij}]
  \end{equation}
  where $\zeta(t,x)$ is scalar perturbation and $\gamma_{ij}(x,t)$ is tensor perturbation which
 is transverse and traceless satisfying,
  $\gamma_{ii}=0$ and $ \partial_{j} \gamma_{ij}=0$.
 In terms of $\gamma_{ij}$ the graviton Lagrangian \eqref{eq1} can be written as,
  \begin{multline}\label{eq3}
  S_{3}^{T}=\int d^4x \sqrt{-g}\left[\frac{M_{pl}^2}{8}\left(\dot{\gamma}_{ij}^2-
  \frac{(\partial_k \gamma_{ij})^2}{a^2}\right)-\frac{\bar{M}_{3}^{2}}{8}\dot{\gamma}_{ij}^2 \right.\\
\left. -  \frac{M_{pl}^2}{8}\left(2 \gamma_{ik} \gamma_{jl}-\gamma_{ij}\gamma_{kl} \right)
\frac{\partial_k \partial_l \gamma_{ij}}{a^2}-\frac{\bar{M}_9}{3!} \dot{\gamma}_{ij}  \dot{\gamma}_{jk}  
\dot{\gamma}_{ki}\right]
  \end{multline}
  
  Equation \eqref{eq3} shows the effect of higher order EFT operators where $\delta K_{\mu}^{\nu}\delta K_{\nu}^{\mu}$ 
  term modifies the dispersion relation of tensor fluctuation \cite{crem2} and 
  $\delta K_{\mu}^{\nu}\delta K_{\nu}^{\rho}\delta K_{\rho}^{\mu}$ introduces a new interaction proportional to
   $\bar{M}_9$ which will contribute to bispectrum. 
    From \eqref{eq3} the typical sound speed of tensor perturbation (speed of propagation
    of GW) can readily be defined as
  \begin{equation}
  c_{\gamma}^2= \frac{M_{pl}^2}{M_{pl}^2-\bar{M}_3^2}
  \end{equation}
  Let us reiterate that {\it this is the most general parity conserving third order graviton Lagrangian.}

  %%%%%%%%%%%%%%%%%%%%%%%%%%%%%%%%%%%%%%%%%%%%%%%%%%%%%%%%%%%%%%%%%%%%%%%%%%%%%%%%%%%%%%%%%%
   \section{Model-independent Tensor Bispectrum}
%{\it Model-independent Tensor Bispectrum}  --
  We now decompose the tensor perturbation in Fourier modes as
  \begin{equation}
  \gamma_{ij}(x,t)=\int \frac{d^3 k}{(2 \pi)^3} \left[\epsilon_{ij}^{+}(\bold{k})
   \gamma_{\bold{k}}^{+}(t)+\epsilon_{ij}^{-}(\bold{k}) \gamma_{\bold{k}}^{-}(t)\right] e^{i \bold{k} .\bold{x}}
  \end{equation}
  Here $\epsilon_{ij}^{s}(\bold{k})$ are polarization tensor and $s=(+,-)$ is the helicity index. The polarization tensor obeys
  %\begin{eqnarray}
  %\epsilon_{ii}(\bold{k})^s=\partial_j \epsilon_{ij}(\bold{k})=0\\
  %\epsilon_{ij}(\bold{k})^s \epsilon_{ij}(\bold{k})^{s'}=\delta_{s s'}\\
  %(\epsilon_{ij}^s(\bold{k}))^{*}=\epsilon_{ij}^s(\bold{-k})
  %\end{eqnarray}
  %\begin{eqnarray}
  $\epsilon_{ii}^s(\bold{k})=\partial_j \epsilon_{ij}^s(\bold{k})=0$,
  $\epsilon_{ij}^s (\bold{k}) \epsilon_{ij}^{s'}(\bold{k})=\delta_{s s'}$,
  $(\epsilon_{ij}^s(\bold{k}))^{*}=\epsilon_{ij}^s(\bold{-k})$.
  %\end{eqnarray}

From \eqref{eq3} we get the solution for the mode function as
\begin{equation}
\gamma_k=\frac{H}{M_{pl}}\frac{c_{\gamma}}{(c_{\gamma} k)^{\frac{3}{2}}}
(1+i c_{\gamma} k \tau) e^{-i c_{\gamma} k \tau}
\end{equation}
Thus we can derive the two point function for tensor modes  in the super horizon limit as
\begin{equation}
\left\langle \gamma_k^s \gamma_{k'}^{s'} \right\rangle = \delta_{s s'} (2 \pi)^3 \delta^{(3)}
({\bold{k}-\bold{k'}}) \frac{H^2}{M_{pl}^2 c_{\gamma}}\frac{1}{k^3}
\end{equation}

To calculate the three point function for tensor perturbation we use IN-IN formalism. In this formalism,
 computation of the three point function will depend on the interaction Hamiltonian 
\begin{multline}\label{eq11}
 H_{int}=\int d^3x a^3 \left[ \frac{M_{pl}^2}{8}\left(2 \gamma_{ik} \gamma_{jl}-
 \gamma_{ij}\gamma_{kl} \right)\frac{\partial_k \partial_l \gamma_{ij}}{a^2} \right. \\
\left. +\frac{\bar{M}_9}{3!} \dot{\gamma}_{ij}  \dot{\gamma}_{jk}  \dot{\gamma}_{ki}\right]
\end{multline}
The first term in \eqref{eq11} is the lowest order contribution of EFT (the so-called Einstein term)
and the 
second term is the contribution of higher power EFT operator.

There can be at most eight combinations of three point functions: 
$\left\langle \gamma_{k_1}^+ \gamma_{k_2}^+ \gamma_{k_3}^+ \right\rangle$, 
$\left\langle \gamma_{k_1}^- \gamma_{k_2}^- \gamma_{k_3}^- \right\rangle$,
 $\left\langle \gamma_{k_1}^+ \gamma_{k_2}^+ \gamma_{k_3}^- \right\rangle$ and its two cyclic permutations,
  $\left\langle \gamma_{k_1}^+ \gamma_{k_2}^- \gamma_{k_3}^- \right\rangle$ and its two cyclic permutations.
  However, very few of them are independent as it will be revealed later on.
  %out of them only two are independent for equilateral limit, whereas for squeezed limit the number of independent combinations are three. 
Further,
  as there is no parity violating term we have
%\begin{eqnarray}
$\left\langle \gamma_{k_1}^+ \gamma_{k_2}^+ \gamma_{k_3}^+ \right\rangle=
\left\langle \gamma_{k_1}^- \gamma_{k_2}^- \gamma_{k_3}^- \right\rangle$.
%\\
%\left\langle \gamma_{k_1}^+ \gamma_{k_2}^+ \gamma_{k_3}^- \right\rangle=
%\left\langle \gamma_{k_1}^+ \gamma_{k_2}^- \gamma_{k_3}^- \right\rangle
%\end{eqnarray}

We use IN-IN formalism to evaluate the three point correlator. In this formalism
time ordering is different than that is used in quantum field theory and expectation value 
for any operator $\mathcal{O}(t)$ can be given as,
\begin{multline}
\langle \mathcal{O}(t)\rangle=
 \left\langle \left(T \exp\left(-i \int_{-\infty}^{t}H_{int}(t')dt'\right)\right)^{\dagger} \mathcal{O}(t)\right.\\
  \left. \left(T \exp\left(-i\int_{-\infty}^{t}H_{int}(t'')dt''\right)\right)\right\rangle
\end{multline}
Where $H_{int}$ is the interaction Hamiltonian.

With the interaction Hamiltonian given by \eqref{eq11}, the correlation functions are given by
\begin{multline}\label{eq14}
\left\langle \gamma_{k_1}^{s_1} \gamma_{k_2}^{s_2} \gamma_{k_3}^{s_3} \right\rangle=
(2 \pi)^3 \delta^{(3)}(\bold{k_1}+\bold{k_2}+\bold{k_3})F(s_1 k_1,s_2 k_2,s_3 k_3)\\
\left(\frac{64 H^4}{c_{\gamma}^2 M_{pl}^4}\frac{A(k_1,k_2,k_3)(s_1  k_1+s_2  k_2+s_3 k_3)^2}{k_1^3 k_2^3 k_3^3}\right. \\
\left. + \frac{4 \bar{M}_9 H^5}{M_{pl}^6}\frac{1}{k_1 k_2 k_3} \frac{1}{(k_1+ k_2+ k_3)^3}\right)
\end{multline}
where
%\begin{equation}
$A(k_1,k_2,k_3)=\frac{K}{32}\left(1-\frac{1}{k^3} \sum_{i \neq j} k_i^2 k_j -\frac{4 k_1 k_2 k_3}{K^3}  \right)$
%\end{equation}
with $\bold{K}=k_1+k_2+k_3$, and
%\begin{equation}
$F(x,y,z)=-\frac{1}{64 x^2 y^2 z^2}(x+y+z)^3 (x+y-z) (x-y+z) (y+z-x).$
%\end{equation}
The term $F(x,y,z)$ is the result of contraction between the polarization tensors \cite{soda}.

From equation \eqref{eq14} we find that the lowest order EFT contribution to bispectrum depends on the 
tensor sound speed $c_{\gamma}$ (or, equivalently, on the EFT parameter $\bar M_3$)
 whereas the size of the higher power EFT term  is solely measured by the parameter $\bar{M}_9$. 
 %As we will justify later on, large non-Gaussianity can, in principle, be obtained  from the lowest order EFT contribution 
 %and any effect of the higher derivative EFT term is, in general, sub-dominant.
 %However, keeping them in the theory can, in practice, include the spectrum of possible inflationary models {\it per se}
 % as subsets. So, for the sake of generality we take into account the effects of all contributions upto third order EFT.

\begin{figure}
	\includegraphics[width=7cm,height=12cm,keepaspectratio]{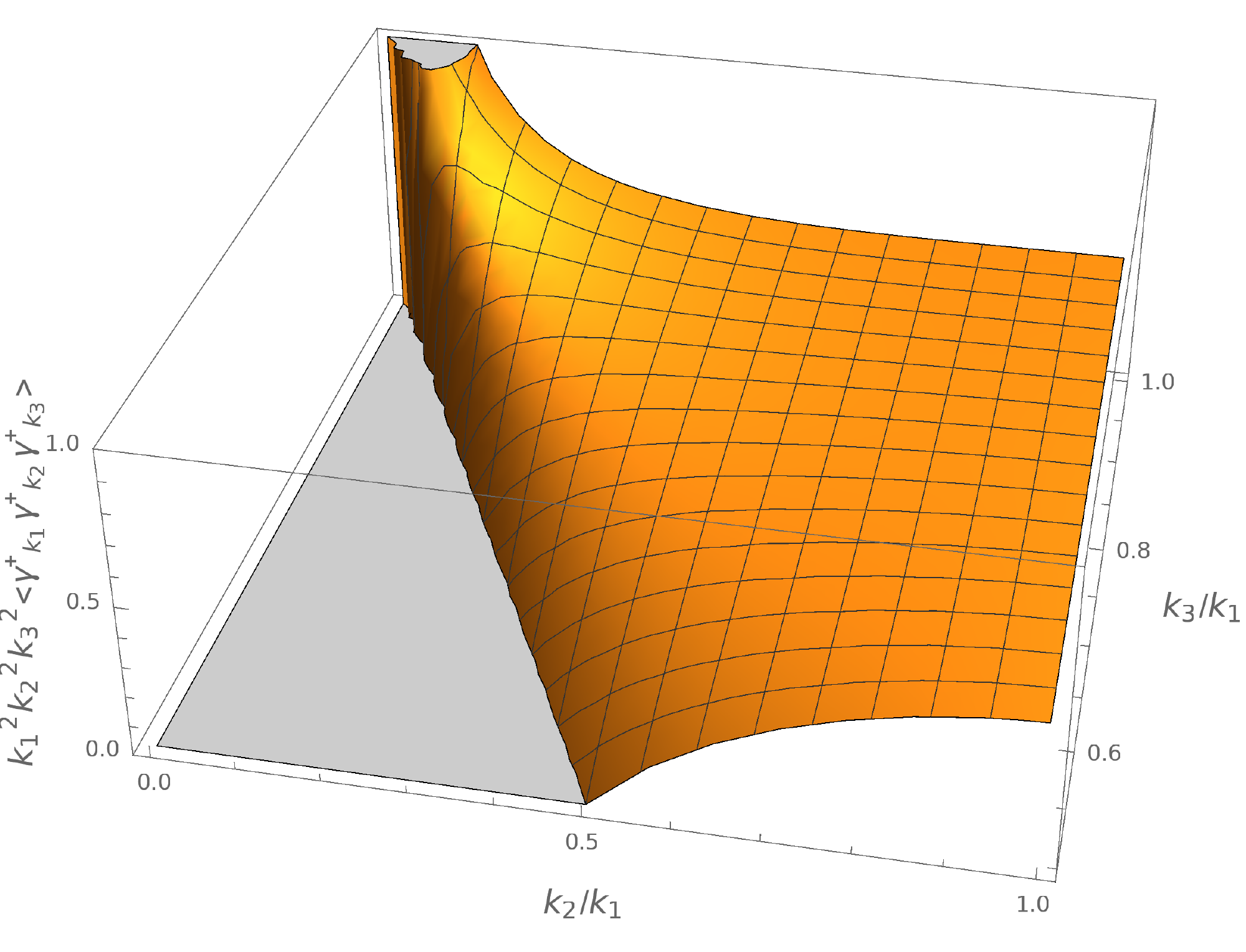}% Here is how to import EPS art
	\caption{The amplitude of bispectrum due to the lowest order EFT term as a function of $k_2/k_1$ and $k_3/k_1$}
	\label{fig1}
\end{figure}

\begin{figure}
	\includegraphics[width=7cm,height=12cm,keepaspectratio]{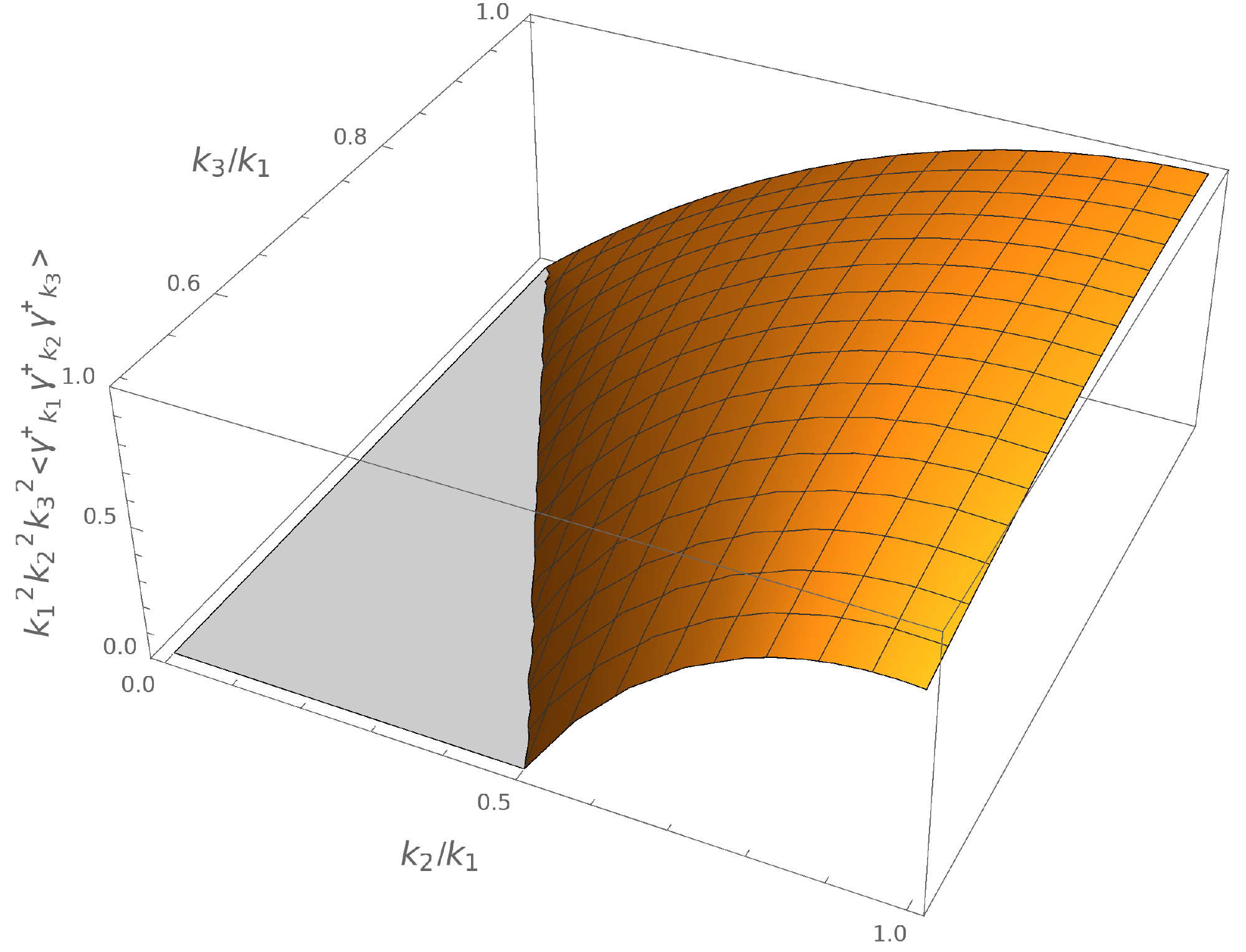}% Here is how to import EPS art
	\caption{The amplitude of bispectrum due to higher power EFT term as a function of $k_2/k_1$ and $k_3/k_1$}
	\label{fig2}
\end{figure}

In figures \ref{fig1} and \ref{fig2}  the amplitude of bispectrum for $\langle\gamma^+_{k_1}\gamma^+_{k_1}\gamma^+_{k_1}\rangle$ helicity combination
(for lowest order and higher power EFT terms respectively)  have been plotted 
as a function of $k_2/k_1$ and $k_3/k_1$. Figure \ref{fig1} shows that the bispectrum  
has a peak in the squeezed limit and  figure \ref{fig2} has a peak in equilateral limit. Here we have only plotted the figures for $\langle\gamma^+_{k_1}\gamma^+_{k_1}\gamma^+_{k_1}\rangle$ as this helicity combination has the largest amplitude and the amplitudes of other helicity combinations are either equal or less than  $\langle\gamma^+_{k_1}\gamma^+_{k_1}\gamma^+_{k_1}\rangle$ combination as shown in the next section. The term $F(x,y,z)$ is the reason that the lowest order EFT term is getting peaked at squeezed limit and the higher order EFT term is peaked at equilateral limit and intermediate limits are becoming sub dominant. As an example the expression of $F(k_1,k_2,k_3)$ reveals that for folded limit where $\frac{k_2}{2}+\frac{k_3}{2}=k_1$, $F(k_1,k_2,k_3)$ is zero for the presence of $(k_2+k_3-k_1)$ term. 

%As it appears,
%though the higher derivative EFT contribution is small compared to the leading order
%term, their functional behaviors are completely different from each other.
%Hence, as argued earlier, both the effects need to be explored in order to keep the spectrum of detection wide open.
As we will justify later on, the lowest order term  has the dominant contribution 
to the bispectrum.
%From the figures, it transpires that tensor bispectra for squeezed limit and
% equilateral limit are our point of interest. We shall thus attempt to 
% investigate for these two limits only.
Therefore,
chances of detecting squeezed limit tensor bispectrum are fairly higher than   equilateral limit.
The figures also reveal that only these two limits should be our point of interest while searching for
tensor non-Gaussianities in future CMB missions.

%%%%%%%%%%%%%%%%%%%%%%%%%%%%%%%%%%%%%%%%%%%%%%%%%%%%%%%%%%%%%%%%%%%%%%%%%%%%%
\section{Templates and Detectability} 
%As is well-known, the usual templates for scalar bispetrum are given by the equilateral and squeezed limits (for a comprehensive
%review, see, for example, \cite{Komatsu:2002db}).
%In the same vein, we try to find out if templates for tensor bispetrum can be obtained in the similar limiting configurations.
%To arrive at the templates for tensor bispetrum, we first obtain the equilateral and squeezed limits of these 
Let us first obtain the correlators for tensor modes.
The equilateral  limit of the correlators look
\begin{multline}
\left\langle \gamma_{k_1}^+ \gamma_{k_2}^+ \gamma_{k_3}^+ \right\rangle|_{eq}=
\left\langle \gamma_{k_1}^- \gamma_{k_2}^- \gamma_{k_3}^- \right\rangle|_{eq}=\\
(2 \pi)^3 \delta^{(3)}(\bold{k_1}+\bold{k_2}+\bold{k_3})\frac{H^4}{16 c_{\gamma}^2 M_{pl}^4}
\left( \frac{459}{2}-\frac{\bar{M}_9 H}{M_{pl}^2} c_{\gamma}^2 \right)\frac{1}{k_1^6}, \\
%\end{multline}
%\begin{multline}
 \left\langle \gamma_{k_1}^+ \gamma_{k_2}^+ \gamma_{k_3}^- \right\rangle|_{eq}= 
\left\langle \gamma_{k_1}^+ \gamma_{k_2}^- \gamma_{k_3}^- \right\rangle|_{eq}= {\rm cyclic ~ perms}=\\
(2 \pi)^3 \delta^{(3)}(\bold{k_1}+\bold{k_2}+\bold{k_3})
 \frac{H^4}{c_{\gamma}^2 M_{pl}^4} \left(\frac{17}{96}-\frac{\bar{M}_9 H}{144 M_{pl}^2} c_{\gamma}^2 \right)
 \frac{1}{k_1^6}
\end{multline}
whereas in the squeezed limit, they take the form
\begin{multline}
\left\langle \gamma_{k_1}^+ \gamma_{k_2}^+ \gamma_{k_3}^+ \right\rangle|_{sq}=
\left\langle \gamma_{k_1}^- \gamma_{k_2}^- \gamma_{k_3}^- \right\rangle|_{sq}=\\
(2 \pi)^3 \delta^{(3)}(\bold{k_1}+\bold{k_2}+\bold{k_3})
\frac{H^4}{ c_{\gamma}^2 M_{pl}^4}\frac{1}{k_1 k_2^3}
 \left(\frac{3}{ k_1^2}-\frac{\bar{M}_9 H}{8 M_{pl}^2} c_{\gamma}^2\frac{1}{k_2^2} \right),\\
%\end{multline}
%\begin{multline}
\left\langle \gamma_{k_1}^+ \gamma_{k_2}^+ \gamma_{k_3}^- \right\rangle|_{sq}=
\left\langle \gamma_{k_1}^+ \gamma_{k_2}^- \gamma_{k_3}^+ \right\rangle|_{sq}=
\left\langle \gamma_{k_1}^- \gamma_{k_2}^- \gamma_{k_3}^+ \right\rangle|_{sq}=\\
\left\langle \gamma_{k_1}^- \gamma_{k_2}^+ \gamma_{k_3}^- \right\rangle|_{sq}=(2 \pi)^3 \delta^{(3)}(\bold{k_1}+\bold{k_2}+\bold{k_3})\\
 \frac{H^4}{16 c_{\gamma}^2 M_{pl}^4} \left(3 -\frac{\bar{M}_9 H}{2 M_{pl}^2}
   c_{\gamma}^2 \right) \frac{k_1}{k_2^7},\\
%\end{multline}
%\begin{multline}
\left\langle \gamma_{k_1}^+ \gamma_{k_2}^- \gamma_{k_3}^- \right\rangle|_{sq}=
\left\langle \gamma_{k_1}^- \gamma_{k_2}^+ \gamma_{k_3}^+ \right\rangle|_{sq}=\\
(2 \pi)^3 \delta^{(3)}(\bold{k_1}+\bold{k_2}+\bold{k_3})
 \frac{H^4}{c_{\gamma}^2 M_{pl}^4}\frac{1}{k_1 k_2^3} \left(\frac{3}{k_1^2}-\frac{\bar{M}_9 H}{8M_{pl}^2}
   c_{\gamma}^2 \frac{1}{k_2^2} \right) 
\end{multline}

We can now define  the dimensionless ``nonlinearity parameter''  in the equilateral limit as
$f_{NL}^T\sim \frac{\left\langle \gamma \gamma \gamma \right\rangle}{\frac{18}{5}
[\frac{2 \pi^2}{k^3} {\cal P}_{\zeta}(k)]^2}$
\cite{planck2}; where ${\cal P}_{\zeta}$ is the dimensionless scalar power spectrum. 
 Using ${\cal P}_{\zeta}(k)=\frac{H^2}{8 \pi^2 M_{pl}^2 c_s \epsilon} (\frac{k}{k_*})^{n_s-1}$ we 
 end up at 
 two independent nonlinearity parameters in equilateral limit:
 \begin{multline}\label{eq21}
f_{NL}^{+++}|_{eq}=f_{NL}^{---}|_{eq}=
\frac{5}{18} \left(\frac{c_s \epsilon}{c_{\gamma}}\right)^2 \left(\frac{459}{2}-\frac{\bar{M}_9 H}{M_{pl}^2}c_{\gamma}^2\right)\\
\left(\frac{k_1}{k_{*}}\right) ^{-2(n_s-1)}
\end{multline}
\begin{multline}\label{eq22}
f_{NL}^{++-}|_{eq}=f_{NL}^{+--}|_{eq}={\rm cyclic ~perms}= \\
\frac{40}{9}\left( \frac{c_s \epsilon}{c_{\gamma}}\right)^2
 \left(\frac{17}{96}-\frac{\bar{M}_9 H}{144 M_{pl}^2}c_{\gamma}^2 \right)
 \left(\frac{k_1}{k_{*}}\right) ^{-2(n_s-1)}
\end{multline}
where the scale $k_1$ is the pivot scale used in the  estimation of all cosmological parameters,
and hence in estimating tensor $f_{NL}$.

%For squeezed limit  $k_1<<k_2=k_3$, 
% $k_1 \sim H$ and we have to measure $f_{NL}^{s_1 s_2 s_3}|_{sq}$ at $k_2=k_3=aH$.
For, squeezed limit wavenumber associated with the squeezed momentum has to be smaller than Hubble parameter. Here we assume the same definition of squeezed limit nonlinearity parameter as equilateral limit 
i.e., $f_{NL}^{s_1 s_2 s_3}|_{sq}\sim \frac{\left\langle \gamma_{k_1}^{s_1} 
\gamma_{k_2}^{s_2} \gamma_{k_3}^{s_3} \right\rangle|_{sq}}{\frac{18}{5}[\frac{2 \pi^2}{k^3} {\cal P}_{\zeta}(k_2)]^2}$
holds good,
at least at the first go. 
Consequently, in squeezed limit we have three  independent nonlinearity parameters:
\begin{multline}\label{eq23}
f_{NL}^{+++}|_{sq}=f_{NL}^{---}|_{sq}=\\
\frac{40}{9}\left( \frac{c_s \epsilon}{c_{\gamma}}\right)^2 \frac{k_2^3}{k_1} \left(\frac{3}{ k_1^2}-
\frac{\bar{M}_9 H}{8 M_{pl}^2} c_{\gamma}^2 \frac{1}{k_2^2}\right)
\left(\frac{k_2}{k_{*}}\right) ^{-2(n_s-1)}
\end{multline}
\begin{multline}\label{eq24}
f_{NL}^{++-}|_{sq}=f_{NL}^{+-+}|_{sq}=f_{NL}^{--+}|_{sq}=f_{NL}^{-+-}|_{sq}=\\
 \frac{5}{18}\left( \frac{c_s \epsilon}{c_{\gamma}}\right)^2
\left(3-\frac{\bar{M}_9 H}{2 M_{pl}^2}
   c_{\gamma}^2 \right) \frac{k_1}{k_2}
 \left(\frac{k_2}{k_{*}}\right) ^{-2(n_s-1)}
\end{multline}
\begin{multline}\label{eq25}
f_{NL}^{+--}|_{sq}=f_{NL}^{-++}|_{sq}=\\
\frac{40}{9}\left( \frac{c_s \epsilon}{c_{\gamma}}\right)^2 \frac{k_2^3}{k_1} \left(\frac{3}{ k_1^2}-
\frac{\bar{M}_9 H}{8 M_{pl}^2} c_{\gamma}^2 \frac{1}{k_2^2}\right)
\left(\frac{k_2}{k_{*}}\right) ^{-2(n_s-1)}
\end{multline}

Equations \eqref{eq21} - \eqref{eq25} 
are generic, model-independent expressions for all tensor nonlinearity parameters $f_{NL}^T$
of our interest. 
%They depend only on couple of EFT parameters.
%So, for any given inflationary model if one can relate the parameters 
 %of the model to EFT parameter $f_{NL}^T$ can be derived for it.
 In this model-independent analysis, along with the usual inflationary parameters
 (scalar sound speed,
 scalar spectral index and first slow roll parameter),
 $c_{\gamma}$ (or, equivalently, $\bar M_3$) and $\bar{M}_9$ are the only two  parameters
 that control the numerical value of $f_{NL}^T$.
 The ratio between the higher order and lowest order EFT contribution is $\mathcal{O}\left(\frac{\bar{M}_9 H}{M_{pl}^2}\right)$.
  But even if $\bar{M}_9$ is as large as $M_{pl}$ 
  the ratio of the order of $\frac{H}{M_{pl}}$. The well-known
  bounds on H on pivot scale being $(\frac{H_*}{M_{pl}})<3.6 \times 10^{-5}$ \cite{planck}, so the contribution of higher order EFT operator with respect to the lowest order EFT operator is less than $10^{-5}$ so it can be concluded that
  $f_{NL}^T$ is practically insensitive to $\bar{M}_9$. Also, $c_{\gamma}^2 \leq 1$
 anyway. 
 One can also set it to unity by a disformal transformation \cite{crem}
that results in a modified Hubble parameter.
 Thus, any effect of 
 the higher derivative EFT term, that mostly reflects the equilateral limit, is sub-dominant.
 Therefore, the lowest order EFT term, that has a peak in the squeezed limit, has the dominant contribution
 to the tensor nonlinearity parameters $f_{NL}^T$.
 %So, considerably large tensor non-Gaussianities could have, in principle, been generated from the lowest order EFT contribution.
 However, there is an overall factor $(c_s \epsilon/c_{\gamma})^2$ outside, that is nothing
 but the tensor-to-scalar ratio ($r$) squared for single field inflation models, and puts additional constraints $r<0.064$ \cite{Akrami:2018odb}. A very rough estimate of the $f_{NL}^T$ for the $\langle \gamma^+_{k_1} \gamma^+_{k_2} \gamma^+_{k_3} \rangle$ combination can be given as, $f_{NL}^{+++}|_{eq}\sim 0.269$ and $f_{NL}^{+++}|_{sq}\sim 56.33$ for $\frac{k_2}{k_1}\sim 10$. 
 %One needs to keep this in mind while investigating for tensor non-Gaussianities by future CMB missions.
 %So, it is nearly impossible to generate large tensor non-gaussianities at least for source-free case.
 If the future observations are sensitive enough and are going to detect tensor non-Gaussianity some day,
 it is expected that they will detect squeezed limit
 tensor bispectrum first.

  The bottomline is that, 
   only equilateral and  squeezed limits should be our point of interest while searching for
tensor non-Gaussianities in future CMB missions as two different operators have peaks in these two limits. Though the contribution of the higher power EFT operator is much smaller than the lowest order EFT operator, any chances of getting the signature of the higher power operator is at equilateral limit. This should be investigated further 
for analysis and comparison with the sensitivity of upcoming CMB missions.
 Nevertheless, 
chances of detecting squeezed limit tensor bispectrum are fairly higher than 
  equilateral limit.
  We believe these two aspects can indeed serve as  motivation to investigate for 
  the non-Gaussian effects
  of primordial GW 
   in future CMB missions using the generic templates proposed in this article.

 %%%%%%%%%%%%%%%%%%%%%%%%%%%%%%%%%%%%%%%%%%%%%%%%%%%%%%%%%%%%%%%%%%%%%%%%%%%%%%%%
%\section{Conclusion}

\section{Conclusions}
%In this $Letter$ we compute the expression of tensor bispectrum using EFT. We showed that in the EFT framework large
% non-Gaussianity can only be achieved with very small sound speed of tensor perturbation. 
% The contribution of higher derivative term is very small. We derive model independent equilateral parameter $f_{NL}^T$ using EFT.
 In summary, let us highlight the major developments made in this article:
\begin{itemize}
  \item Developed a model-independent framework for calculating tensor non-Gaussianities
 based  on EFT of inflation. 
 \item Proposed generic templates for  tensor non-Gaussianities that may be useful
 for future CMB missions.
% \item Discussed prospects of detecting tensor non-Gaussianities in future CMB missions
 %for primordial gravitational waves 
 %via their non-Gaussian effects
 %tensor modes via large non-Gaussianities 
 %by tuning the speed of propagation of primordial gravitational  waves.
 \item Discussed why only equilateral and  squeezed limits 
 should be our point of interest while searching for
tensor non-Gaussianities in future CMB missions.
  \item Found that chances of detecting squeezed limit tensor bispectrum are fairly higher than 
  equilateral limit.
%\item The analysis is based solely on EFT of inflation and the results are quite
%generic with very few tuning parameters.
\end{itemize}

 {\it Acknowledgments} -- We thank Eiichiro Komatsu and Aniket Agrawal for early discussions.
 We also thank Paolo Creminelli and Filippo Vernizzi for useful comments.
 AN thanks ISI Kolkata for financial support through Senior Research Fellowship.


\begin{thebibliography}{}

\bibitem{ligo1} B. P. Abbott {\it et al}. [LIGO Scientific Collaboration and Virgo Collaboration],
Phys. Rev. Lett. {\bf 116}, 061102 (2016)

\bibitem{ligo2} B. P. Abbott {\it et al.} [LIGO Scientific Collaboration and Virgo Collaboration],
Phys. Rev. Lett. {\bf 116}, 241103 (2016)

\bibitem{ligo3} B. P. Abbott {\it et al.}
[LIGO Scientific Collaboration and Virgo Collaboration],
Phys. Rev. Lett. {\bf 119}, 161101 (2017)



\bibitem{planck} 
P.~A.~R.~Ade {\it et al.} [Planck Collaboration],
  %``Planck 2015 results. XX. Constraints on inflation,''
  Astron.\ Astrophys.\  {\bf 594}, A20 (2016)
  %doi:10.1051/0004-6361/201525898
  %[arXiv:1502.02114 [astro-ph.CO]].
  %%CITATION = doi:10.1051/0004-6361/201525898;%%
  %1650 citations counted in INSPIRE as of 11 Jun 2018

\bibitem{Akrami:2018odb} 
  Y.~Akrami {\it et al.} [Planck Collaboration],
  %``Planck 2018 results. X. Constraints on inflation,''
  arXiv:1807.06211 [astro-ph.CO].  


\bibitem{planck2}  
  P.~A.~R.~Ade {\it et al.} [Planck Collaboration],
  %``Planck 2015 results. XVII. Constraints on primordial non-Gaussianity,''
  Astron.\ Astrophys.\  {\bf 594}, A17 (2016)
 % doi:10.1051/0004-6361/201525836
  %[arXiv:1502.01592 [astro-ph.CO]].
  %%CITATION = doi:10.1051/0004-6361/201525836;%%
  %481 citations counted in INSPIRE as of 11 Jun 2018
  
   
  \bibitem{Shiraishi:2014ila} 
  M.~Shiraishi, M.~Liguori and J.~R.~Fergusson,
  %``Observed parity-odd CMB temperature bispectrum,''
  JCAP {\bf 1501},  007 (2015)
  %doi:10.1088/1475-7516/2015/01/007
  %[arXiv:1409.0265 [astro-ph.CO]].
  %%CITATION = doi:10.1088/1475-7516/2015/01/007;%%
  %16 citations counted in INSPIRE as of 11 Jun 2018
  
  \bibitem{COrE} 
C. Armitage-Caplan {\it et al.},  [COrE Collaboration],
arXiv:1102.2181 [astro-ph.CO]
  
\bibitem{Matsumura:2013aja} 
  T.~Matsumura {\it et al.}, [LiteBIRD Collaboration],
  %``Mission design of LiteBIRD,''
  J.\ Low.\ Temp.\ Phys.\  {\bf 176}, 733 (2014)
 % doi:10.1007/s10909-013-0996-1
  %[arXiv:1311.2847 [astro-ph.IM]].
  
\bibitem{Suzuki:2018cuy} 
  A.~Suzuki {\it et al.}, [LiteBIRD Collaboration],
  %``The LiteBIRD Satellite Mission - Sub-Kelvin Instrument,''
  J.\ Low.\ Temp.\ Phys.\  (2018) https://doi.org/10.1007/s10909-018-1947-7
  %[arXiv:1801.06987 [astro-ph.IM]].  
  
\bibitem{Abazajian:2016yjj} 
  K.~N.~Abazajian {\it et al.} [CMB-S4 Collaboration],
  %``CMB-S4 Science Book, First Edition,''
  arXiv:1610.02743 [astro-ph.CO] 
  
\bibitem{Andre:2013nfa} 
  P.~Andr\'{e} {\it et al.} [PRISM Collaboration],
  %``PRISM (Polarized Radiation Imaging and Spectroscopy Mission): An Extended White Paper,''
  JCAP {\bf 1402}, 006 (2014)
 % doi:10.1088/1475-7516/2014/02/006
  %[arXiv:1310.1554 [astro-ph.CO]].  

\bibitem{Kogut:2011xw} 
  A.~Kogut {\it et al.}, [PIXIE Collaboration],
  %``The Primordial Inflation Explorer (PIXIE): A Nulling Polarimeter for Cosmic Microwave Background Observations,''
  JCAP {\bf 1107}, 025 (2011)
%  doi:10.1088/1475-7516/2011/07/025
  %[arXiv:1105.2044 [astro-ph.CO]].

%\cite{Bartolo:2018qqn}
\bibitem{Bartolo:2018qqn} 
  N.~Bartolo {\it et al.},
  %``Probing non-Gaussian Stochastic Gravitational Wave Backgrounds with LISA,''
  arXiv:1806.02819 [astro-ph.CO].
 

\bibitem{mald1} 
  J.~M.~Maldacena,
  %``Non-Gaussian features of primordial fluctuations in single field inflationary models,''
  JHEP {\bf 0305}, 013 (2003)
%  doi:10.1088/1126-6708/2003/05/013
  %[astro-ph/0210603].

\bibitem{mald2} 
  J.~M.~Maldacena and G.~L.~Pimentel,
  %``On graviton non-Gaussianities during inflation,''
  JHEP {\bf 1109}, 045 (2011)
%  doi:10.1007/JHEP09(2011)045
  %[arXiv:1104.2846 [hep-th]].
  
 
\bibitem{soda}  J.~Soda, H.~Kodama and M.~Nozawa,
  %``Parity Violation in Graviton Non-gaussianity,''
  JHEP {\bf 1108}, 067 (2011)
  %doi:10.1007/JHEP08(2011)067
  %[arXiv:1106.3228 [hep-th]].
  

\bibitem{yamaguchi} 
  X.~Gao, T.~Kobayashi, M.~Yamaguchi and J.~Yokoyama,
  %``Primordial non-Gaussianities of gravitational waves in the most general single-field inflation model,''
  Phys.\ Rev.\ Lett.\  {\bf 107}, 211301 (2011)
%  doi:10.1103/PhysRevLett.107.211301
  %[arXiv:1108.3513 [astro-ph.CO]].
  
  %\cite{Kobayashi:2011nu}
\bibitem{Kobayashi:2011nu} 
  T.~Kobayashi, M.~Yamaguchi and J.~Yokoyama,
  %``Generalized G-inflation: Inflation with the most general second-order field equations,''
  Prog.\ Theor.\ Phys.\  {\bf 126}, 511 (2011)
  %doi:10.1143/PTP.126.511
  %[arXiv:1105.5723 [hep-th]].
  %%CITATION = doi:10.1143/PTP.126.511;%%
  %486 citations counted in INSPIRE as of 25 Jun 2018


\bibitem{sorbo} 
  J.~L.~Cook and L.~Sorbo,
  %``An inflationary model with small scalar and large tensor nongaussianities,''
  JCAP {\bf 1311}, 047 (2013)
%  doi:10.1088/1475-7516/2013/11/047
  %[arXiv:1307.7077 [astro-ph.CO]].

\bibitem{aniket} 
  A.~Agrawal, T.~Fujita and E.~Komatsu,
  %``Large Tensor Non-Gaussianity from Axion-Gauge Fields Dynamics,''
  Phys.\ Rev.\ D {\bf 97},  103526 (2018)
%  doi:10.1103/PhysRevD.97.103526
  %[arXiv:1707.03023 [astro-ph.CO]].

\bibitem{crem2} 
  C.~Cheung, P.~Creminelli, A.~L.~Fitzpatrick, J.~Kaplan and L.~Senatore,
  %``The Effective Field Theory of Inflation,''
  JHEP {\bf 0803}, 014 (2008)
%  doi:10.1088/1126-6708/2008/03/014
  %[arXiv:0709.0293 [hep-th]].

\bibitem{weinberg} 
  S.~Weinberg,
  %``Effective Field Theory for Inflation,''
  Phys.\ Rev.\ D {\bf 77}, 123541 (2008)
%  doi:10.1103/PhysRevD.77.123541
 % [arXiv:0804.4291 [hep-th]].
  
 % \bibitem{Komatsu:2002db} 
 % E.~Komatsu,
 % ``The pursuit of non-gaussian fluctuations in the cosmic microwave background,''
 % arXiv:astro-ph/0206039
  %%CITATION = ASTRO-PH/0206039;%%
  %97 citations counted in INSPIRE as of 14 Jun 2018
  
%  \bibitem{Guzzetti:2016mkm} 
%  M.~C.~Guzzetti, N.~Bartolo, M.~Liguori and S.~Matarrese,
  %``Gravitational waves from inflation,''
 % Riv.\ Nuovo Cim.\  {\bf 39}, no. 9, 399 (2016)
  %doi:10.1393/ncr/i2016-10127-1
  %arXiv:1605.01615 [astro-ph.CO]
  
  %\cite{Cai:2015yza}
%\bibitem{Cai:2015yza} 
 % Y.~Cai, Y.~T.~Wang and Y.~S.~Piao,
  %``Is there an effect of a nontrivial $c_T$ during inflation?,''
  %Phys.\ Rev.\ D {\bf 93}, no. 6, 063005 (2016)
  %doi:10.1103/PhysRevD.93.063005
  %[arXiv:1510.08716 [astro-ph.CO]].
  %%CITATION = doi:10.1103/PhysRevD.93.063005;%%
  %19 citations counted in INSPIRE as of 25 Jun 2018
  
  %\cite{Cai:2016ldn}
%\bibitem{Cai:2016ldn} 
 % Y.~Cai, Y.~T.~Wang and Y.~S.~Piao,
 % %``Propagating speed of primordial gravitational waves and inflation,''
 % Phys.\ Rev.\ D {\bf 94}, no. 4, 043002 (2016)
  %doi:10.1103/PhysRevD.94.043002
  %[arXiv:1602.05431 [astro-ph.CO]].
  %%CITATION = doi:10.1103/PhysRevD.94.043002;%%
  %14 citations counted in INSPIRE as of 25 Jun 2018
  
\bibitem{crem} 
 P.~Creminelli, J.~Gleyzes, J.~Noreña and F.~Vernizzi,
 % ``Resilience of the standard predictions for primordial tensor modes,''
  Phys.\ Rev.\ Lett.\  {\bf 113},  231301 (2014)
%  doi:10.1103/PhysRevLett.113.231301
  %[arXiv:1407.8439 [astro-ph.CO]].
  


\end{thebibliography}

\end{document}